\documentclass[%
 aps,
 pre,
 amsmath,amssymb,
 reprint,%
]{revtex4-1}

\usepackage{graphicx}
\usepackage{dcolumn}
\usepackage{bm}

\usepackage[utf8]{inputenc}
\usepackage[T1]{fontenc}
\usepackage{mathptmx}
\usepackage{etoolbox}
\usepackage{xcolor}

\makeatletter
\def\@email#1#2{%
 \endgroup
 \patchcmd{\titleblock@produce}
  {\frontmatter@RRAPformat}
  {\frontmatter@RRAPformat{\produce@RRAP{*#1\href{mailto:#2}{#2}}}\frontmatter@RRAPformat}
  {}{}
}%
\makeatother
\begin{document}

\preprint{AIP/123-QED}

\title{Dissecting mode-coupling theory for supercooled liquids}
\author{Ilian Pihlajamaa}
\affiliation{%
Soft Matter \& Biological Physics, Department of Applied Physics,
Eindhoven University of Technology, P.O. Box 513, 5600MB Eindhoven, The Netherlands
}%
\author{Liesbeth M.C. Janssen}%
 \email{l.m.c.janssen@tue.nl}
\affiliation{%
Soft Matter \& Biological Physics, Department of Applied Physics,
Eindhoven University of Technology, P.O. Box 513, 5600MB Eindhoven, The Netherlands
}%

\date{\today}

\begin{abstract}
The mode-coupling theory (MCT) of the glass transition ranks among the most successful first-principles kinetic theories to describe glassy dynamics. 
However, MCT does not fully account for crucial aspects of the dynamics near the glass transition. To facilitate improving the theory, we critically test 
the approximations inherent in MCT for a supercooled mixture using Brownian dynamics simulations. Although each MCT approximation significantly impacts the predicted dynamics, our findings show that long-time cancellations of errors occur due to static and dynamic approximations of four-point correlation functions, with the validity of these approximations remaining relatively constant across different temperatures. Notably, the MCT form of the memory functional maintains remarkably high accuracy even in the supercooled regime when evaluated with the intermediate scattering function from simulations. The primary discrepancies between theoretical predictions and experimental results arise from the self-consistent nature of the MCT equations, which amplify minor errors in the memory kernel. This suggests that minor corrections to the memory functional could substantially enhance the theory's predictive accuracy.
\end{abstract}

\maketitle

\newcommand{\kk}[1]{\textbf{k}_#1}
\newcommand{\rk}[1]{\rho_{\textbf{k}_#1}}
\newcommand{\pp}{\mathcal{P}}
\newcommand{\qq}{{\mathcal{Q}}}
\newcommand{\ket}[1]{|#1\rangle}
\newcommand{\bra}[1]{\langle#1|}

\newcommand{\cors}[2]{\langle#1 | #2\rangle}
\newcommand{\cort}[3]{\langle#1 | #2 | #3\rangle}
\newcommand{\corte}[2]{\langle#1 | e^{\varOmega t}| #2\rangle}

\section{Introduction}

 The glass transition is an important phenomenon in condensed matter physics, marking the transition of a liquid into an amorphous solid state upon supercooling below the freezing temperature \cite{ediger1996supercooled, angell2000relaxation, berthier2011theoretical, cavagna2009supercooled, debenedetti2001supercooled, lubchenko2007theory}. Unlike crystalline solids, which exhibit a sharp transition and long-range order, glasses form with no clear structural change across the transition. During this process, the molecular dynamics slow down while the viscosity and relaxation time increase dramatically. Below the glass transition temperature $T_g$, the viscosity becomes so high that the material appears solid on any reasonable timescale.

Mode-coupling Theory (MCT), developed in the 1980s, is a prominent theoretical framework for describing the dynamics of supercooled liquids and the glass transition \cite{bengtzelius1984dynamics, leutheusser1984dynamical, gotze2009complex,das2004mode, reichman2005mode, janssen2018mode, nagele1999cooperative}. MCT provides a microscopic, first-principles approach, predicting the slowing down of dynamics and identifying a critical temperature $T_c$ below which the system becomes non-ergodic. Through a series of approximations, the theory arrives at a self-consistent equation of motion for the auto-correlation function of the microscopic density field, representing the only slow variable of interest in the theory. In the trivial low-density limit, this auto-correlation function, known as the intermediate scattering function, decays as a damped harmonic oscillator \cite{nagele1999cooperative} or exponential in Newtonian and Brownian liquids, respectively. The associated decay rate is called the bare frequency. The non-trivial finite-density effects are introduced in the theory non-perturbatively through the inclusion of the so-called memory kernel, which couples the decay of the chosen slow mode (of the density field) to the remainder of the system. Mode-coupling theory approximates the memory kernel as a bi-linear functional of the intermediate scattering function, creating a closed equation that can be solved self-consistently. 

One of MCT's primary achievements is its fit-parameter-free prediction of the steep increase in relaxation times as liquids approach $T_c$, which qualitatively agrees with observations in the mildly supercooled regime. Additionally, MCT makes several predictions about the dynamical scaling exponents that govern the dynamics of supercooled liquids, many of which have been verified by simulation studies \cite{gotze1999recent, kob1995testing, kob1995testing2, lupi2024mode}. The theory has been successfully applied to many different systems and typically yields good qualitative results compared to computer simulations. Moreover, the theory has been extended to, \textit{e.g.}, systems under shear \cite{fuchs2009mode, brader2010nonlinear}, in confinement \cite{krakoviack2007mode, lang2010glass}, and systems composed of self-propelling particles \cite{szamel2015glassy, szamel2016theory, feng2017mode, liluashvili2017mode, debets2022active, debets2023mode}.  

Despite these successes, MCT also has limitations. Most notably, the theory predicts a power-law divergence of the relaxation time at $T_c$ that is not observed experimentally. While a power-law regime is often observed, it is in many systems superseded by an Arrhenius law \cite{mallamace2010transport, mallamace2013dynamical}. The associated crossover temperature is labeled as $T_{\mathrm{MCT}}$ since it signifies the end of the power-law regime that MCT predicts, and thus the end of the regime in which MCT is thought to be applicable. However, it is important to note that $T_c$ (obtained from fit-parameter free MCT predictions) typically is significantly higher than the fitted $T_\mathrm{MCT}$ (routinely by a factor of 2) \cite{berthier2010critical} which in turn deviates significantly from $T_g$ in many materials. 
Additionally, as a mean-field-like theory, MCT focuses on two-point correlation functions and does not fully account for spatial fluctuations such as activated events, dynamic heterogeneities, and collective structural features, which seem to be critical near the glass transition \cite{berthier2010critical, berthier2011theoretical}. These heterogeneities become more pronounced as the system approaches the glassy state, leading to deviations from MCT predictions. In the context of MCT, such heterogeneities are absent because the theory does not incorporate dynamic (off-diagonal) higher-order correlations between density fluctuations. Instead, these are treated as Gaussian and thus diagonalized and factorized, even though the off-diagonal contributions have been argued to be highly non-Gaussian in the low-$T$ regime \cite{cates2006current}, which was recently born out by simulation results in the static case \cite{pihlajamaa2023emergent}. 

To address these limitations of MCT, there have been several proposed approaches. Restricting ourselves on theories that rely on projection operator techniques in the original spirit of MCT \cite{bengtzelius1984dynamics, leutheusser1984dynamical, gotze2009complex,das2004mode, reichman2005mode, janssen2018mode}, we can essentially classify these attempts into three groups. 
Firstly, one may phenomenologically modify or add terms in the mode-coupling equations of motion or memory kernel to gain increased accuracy compared to experiments and simulations. The simplest example is the introduction of an \textit{ad hoc} rescaling of the density or temperature to achieve with agreement experimental results \cite{banchio1999rheology, voigtmann2003dynamics, flenner2005relaxation, weysser2010structural, berthier2010critical, diaz2024rescaled, amokrane2017glass}. This procedure has yielded quantitative agreement in the mildly supercooled regime. Another example is the additional inclusion of phenomenological resolved variables beyond the density modes \cite{oxtoby1986generalized}. The disadvantage of the approaches in this category is that they rely in some way on phenomenological models or fitting parameters, and therefore lose predictive power for new systems.

The second group comprises theories that try to improve classical MCT by including more or different slow microscopic variables in the construction of the theory. The idea is that if one includes more slow modes explicitly, then their effects are treated in the bare frequency of the theory and thus taken out of the memory kernel, making the computation of the latter simpler or even redundant (in the case that all important variables are treated explicitly). Examples are the inclusion of products of two \cite{liu1997microscopic, manno1999microscopic} or more \cite{liu1997reexamination, wu2005high, szamel2007dynamics,szamel2004gaussian} density modes. While some of these approaches have been shown to avoid the spurious ergodicity-breaking transition of MCT, the resulting equations are often restrictively complex to solve numerically for realistic liquids. One of the first theories in this class is called Extended MCT, which includes certain non-linear current couplings that result in so-called hopping-like terms in the equation of motion \cite{gotze1987glass, das1986fluctuating, schmitz1993absence, mazenko1994density}.  While their correctness is debated \cite{cates2006current}, these hopping terms also remove the non-ergodicity transition. 

The third category consists of approaches that directly attempt to circumvent or avoid approximations made to the memory kernel. These include the explicit inclusion of static triplet correlations to avoid the convolution approximation \cite{bosse1982mode, barrat1989liquid, sciortino2001debye, kob2002quantitative, ayadim2011mode, luo2022many} and the approach called generalized mode-coupling theory which evades the factorization approximation by constructing a full equation of motion for the four-point dynamic density correlation \cite{szamel2003colloidal, wu2005high, mayer2006cooperativity, szamel2008diagrammatic, szamel2013mode, luo2020generalized1, luo2020generalized, biezemans2020glassy, debets2021generalized}. While these approaches also increase the computational complexity of the theory, numerical results are available and have been promising.

Unfortunately, it remains unclear which of the approximations made to the memory kernel of mode-coupling theory is responsible for which of MCT's limitations. This has resulted in a divide not only in the methods by which improvements have been attempted but also in the approximations they address. Additional difficulty arises because the approximations within MCT are to a large extent uncontrolled and unintuitive. However, in prior work, we have disentangled and tested each of the approximations of MCT in a simple equilibrium liquid by explicitly measuring the memory kernel after each approximation in the theory from computer simulations \cite{pihlajamaa2023unveiling}. This has allowed us to show that a class of attempts called ``diagonal generalized mode-coupling theory" is unhelpful, even though they may give quantitative improvements, because they attempt to improve upon approximations that are exact in the thermodynamic limit. For simplicity, the previous work has considered only single-component fluids at temperatures above the melting point, which limited its applicability to the glass transition. Here, we extend this analysis to a multi-component system in the supercooled regime, which for the first time allows us to systematically test the mode-coupling theory approximations in the theory's intended regime of validity. 

The paper is organized as follows. Sections \ref{sec:theory} and \ref{sec:mct} provide a concise derivation of MCT delineating both the exact equations of motion and the various approximations made in the theory, respectively. Section \ref{sec:methods} describes the numerical simulation methods used to study the supercooled liquid. Section \ref{sec:results} presents the results of the numerical validation of MCT's approximations. Finally, Section \ref{sec:conclusion}  discusses the implications of our findings for future efforts to improve MCT.

\section{Exact dynamical equations}\label{sec:theory}

We start by describing the exact equations of motion of the intermediate scattering function $F(\textbf{k},t)$ in a system of Brownian particles. 
Consider a three-dimensional dense mixture of $N$ particles in volume $V$ at temperature $T$. The $N$ particles are divided into $N_s$ distinct species such that $N = \sum_{\alpha=1}^{N_s} N^\alpha$, where $N^\alpha$ is the number of particles of type $\alpha$.  The locations of these particles at time $t$ are denoted by $\textbf{r}^\alpha_j(t)$, for $j=1\ldots N^\alpha$. For convenience, we also define the partial fractions $x^\alpha = N^\alpha/N$ and partial densities $\rho^\alpha = N^\alpha/V$. If the particles do not have internal degrees of freedom, their structural dynamics are encoded in the correlation of density fluctuations over time. In Fourier space, this correlation function is the intermediate scattering function, defined as
\begin{equation}\label{eq:F}
    F^{\alpha\beta}(k,t) = \frac{1}{N}\corte{\rho^\alpha_\textbf{k}}{\rho^\beta_\textbf{k}},
\end{equation}
where we have introduced the density mode $\ket{\rho_\textbf{k}} = \sum_{j=1}^N\exp(i\textbf{k}\cdot \textbf{r}_j)$, at wave vector $\textbf{k}$, the length of which we denote with $k=|\textbf{k}|$. 

The operator $e^{\varOmega t}$ propagates any observable on its right by an amount of time $t$. Here, the Smoluchowski operator $\varOmega$ is defined as
\begin{equation}
    \varOmega = \sum_{\alpha=1}^{N_s}\sum_{i=1}^{N^\alpha} D^\alpha \nabla_i^\alpha \cdot \left(\nabla_i^\alpha - \beta \textbf{F}_i^\alpha\right),
\end{equation}
in which $D^\alpha$ is the diffusion constant of the particles of species $\alpha$, $\beta=1/k_BT$ is the inverse thermal energy, $\nabla_i^\alpha$ is the gradient with respect to $\textbf{r}_i^\alpha$, and $\textbf{F}_i^\alpha=-\nabla_i^\alpha U(\left\{\textbf{r}_i^\alpha\right\})$ is the force acting on particle $\textbf{r}_i^\alpha$. Here we have defined $\left\{\textbf{r}_i^\alpha\right\}$ to be the set of phase space variables, and $U(\left\{\textbf{r}_i^\alpha\right\})$ as the total potential energy at a given configuration.
The ensemble average introduced in Eq.~\eqref{eq:F} should be interpreted as  
\begin{equation}
    \cort{A}{\mathcal{O}}{B} = \int \mathrm{d}\left\{\textbf{r}_i^\alpha\right\} A^\dagger(\left\{\textbf{r}_i^\alpha\right\}) \mathcal{O} \left[B(\left\{\textbf{r}_i^\alpha\right\}) \exp(-\beta U(\left\{\textbf{r}_i^\alpha\right\}))\right].
\end{equation}
for any two observables $\ket{A(\left\{\textbf{r}_i^\alpha\right\})}$ and $\ket{B(\left\{\textbf{r}_i^\alpha\right\})}$ and operator $\mathcal{O}$.

Using the projector operators $\mathcal{P}$, $\mathcal{P}'$ and their complements $\mathcal{Q}$, and $\mathcal{Q}$, defined as
\begin{equation}
    \pp = \ket{\rho_\textbf{k}^\alpha}\cors{\rho_\textbf{k}^{\alpha}}{\rho_\textbf{k}^\beta}^{-1}\bra{\rho_\textbf{k}^\beta},\qquad \qq = 1 - \pp,
\end{equation}

and 

\begin{equation}
    \pp' = \ket{\rho_\textbf{k}^\alpha}\cort{\rho_\textbf{k}^{\alpha}}{\varOmega}{\rho_\textbf{k}^\beta}^{-1}\bra{\rho_\textbf{k}^\beta \varOmega}, \qquad \qq' = 1 - \pp',
\end{equation}
the formalism of Mori and Zwanzig allows us to write an equation of motion for the intermediate scattering function \cite{gotze2009complex, pihlajamaa2023unveiling}
\begin{align}\label{eq:eom}
    \frac{d\textbf{F}}{dt}(k,t) + \textbf{H}(k)\textbf{S}^{-1}(k)\textbf{F}(k,t) + \int \mathrm{d}\tau \textbf{M}(k,t-\tau) \frac{d}{d\tau}\textbf{F}(k,\tau) = 0
\end{align}
where bold-face symbols denote vector or matrix quantities constructed from the species-specific properties $\left((\textbf{X})^{\alpha\beta} = X^{\alpha\beta}\right)$. Here, we have introduced
\begin{align}\label{eq:K}
    M^{\alpha\beta}(k,t) &= \frac{1}{N}\cort{R_\textbf{k}^\alpha}{e^{\qq\varOmega\qq'\qq t}}{R_\textbf{k}^\gamma} \left(\textbf{H}^{-1}\right)^{\gamma\beta}(k), \\
    H^{\alpha\beta}(k) &= -\frac{1}{N}\cort{\rho_\textbf{k}^\alpha}{\varOmega}{\rho_\textbf{k}^\beta} = k^2x^\alpha D^\alpha \delta^{\alpha\beta},\\
    S^{\alpha\beta}(k) &= \frac{1}{N}\cors{\rho_\textbf{k}^\alpha}{\rho_\textbf{k}^\beta}, 
\end{align}
which are respectively called the (exact) memory kernel, the diffusion matrix, and the static structure factor. The matrix product of the diffusion matrix and the inverse static structure factor is also called the bare frequency, as it dictates the decay of $\textbf{F}(k,t)$ in the absence of memory. In the definition of the memory kernel, we have introduced the fluctuating force as $\ket{R_\textbf{k}^\alpha} = \qq\varOmega\ket{\rho_\textbf{k}^\alpha}$, which is orthogonal to the density mode by construction. Note that the time evolution of the memory kernel [Eq.~\eqref{eq:K}], evolves with orthogonal dynamics by virtue of the projection operators in the exponent $e^{\qq\varOmega\qq'\qq t}$. Throughout this work, any repeated Greek index that does not appear on the left-hand side is summed over.

It is possible to express the memory kernel as the solution to a Volterra integral equation of which the other terms evolve with standard dynamics, that is, with operator $e^{\varOmega t}$, which can therefore be measured in simulation \cite{pihlajamaa2023unveiling}. The result is
\begin{equation}\label{eq:volterra}
    \textbf{M}(k,t) = \textbf{M}_\varOmega(k,t) + \int_0^t\mathrm{d}\tau \textbf{M}(k,t-\tau)\textbf{W}(k,\tau)
\end{equation}
where $\textbf{M}_\varOmega(k,t)$ is the correlation function of the fluctuating force with standard dynamics, and $ \textbf{W}^{\alpha\beta}(k,t) = \frac{1}{N}\cort{\rho_\textbf{k}^\alpha\varOmega}{e^{\Omega t}}{R_\textbf{k}^\gamma} \left(\textbf{H}^{-1}\right)^{\gamma\beta}(k)$.

\section{Mode-coupling theory}\label{sec:mct}

While we have derived an exact equation of motion for the intermediate scattering function, Eq.~\eqref{eq:eom}, it is yet of little practical use since the required memory kernel evolves with dynamics unlike that of Brownian particles. This renders it difficult to approximate the kernel in terms of known correlation functions. Therefore, one way forward is to neglect the effect of the orthogonal projectors in the exponent, letting $\exp(\qq\varOmega\qq'\qq t) \to \exp(\varOmega t)$. We denote the resulting memory kernel as $\textbf{M}_\varOmega(k,t)$, which represents our first approximation. 

To proceed with the complete MCT derivation, we recall that the fluctuating force $\ket{R_\textbf{k}^\alpha}$ contains the effects of all interaction forces within our Brownian liquid on the intermediate scattering function. One can show that the fluctuating force can be rewritten in terms of pair densities 
\begin{align}
    \textbf{R}_i^\alpha &=  \qq\varOmega\ket{\rho_\textbf{k}^\alpha} \\&= \qq\left[D^\alpha\sum_{j=1}^{N^\alpha}\left[i\beta (\textbf{k}\cdot\textbf{F}_j^\alpha)-k^2\right]e^{i(\textbf{k}\cdot\textbf{r}^\alpha_j)}\right]\\
    &= D^\alpha\beta\sum_\gamma\int\mathrm{d}\textbf{q}(\textbf{k}\cdot\textbf{q})\hat{u}^{\alpha\gamma}(\textbf{q})\label{eq:R_doublets}\\&\qquad\times\left[ \cors{\rho_\textbf{k}^\mu}{\rho_\textbf{k}^\nu}^{-1}\cors{\rho_\textbf{k}^\nu}{\rho_\textbf{q}^\gamma\rho_{\textbf{k}-\textbf{q}}^\alpha}\ket{\rho^\mu_\textbf{k}} - \ket{\rho_\textbf{q}^\gamma\rho_{\textbf{k}-\textbf{q}}^\alpha}\right],\nonumber
\end{align}
where $\hat{u}^{\alpha\beta}(\textbf{k})$ is the Fourier transform of the pair interaction potential $u^{\alpha\beta}(\textbf{r}) = \int \mathrm{d}\textbf{q}\hat{u}^{\alpha\beta}(\textbf{q})\exp(i\textbf{q}\cdot\textbf{r})$ (if it exists). This means that, without approximations, the fluctuating force can be projected on the space spanned by density pairs. This is in contrast to the Newtonian case, where current couplings also must be included.

While the fluctuating force that evolves in time with reduced dynamics $\left[\exp(\qq\varOmega\qq'\qq t)\ket{R^\alpha_\textbf{k}}\right]$ is orthogonal with respect to the density modes at all $t$, the fluctuating force that evolves with standard dynamics ($\exp(\varOmega t)\ket{R^\alpha_\textbf{k}}$) is not. To account for this, we must project to the subspace of pair modes which is orthogonal with respect to pair singlets. Specifically, we introduce the orthogonalized density doublet
\begin{equation}\label{eq:Q2}
\ket{Q^{\alpha\beta}_{\textbf{q}\textbf{q}'}} =  \qq \ket{\rho^\alpha_\textbf{q}\rho^\beta_{\textbf{q}'}} \approx \ket{\rho^\alpha_\textbf{q}\rho^\beta_{\textbf{q}'}} - \frac{1}{\left(x^\gamma\right)^2}\ket{\rho^\gamma_{\textbf{q}+\textbf{q}'}}S^{\alpha\gamma}(q)S^{\gamma\beta}(q').
\end{equation}
Here we have used the convolution approximation
\begin{align}    \cors{\rho^\alpha_{\textbf{k}_1+\textbf{k}_2}}{&\rho^\beta_{\textbf{k}_1}\rho^\gamma_{\textbf{k}_2}} \\&\approx \frac{1}{(N^\mu)^2}\cors{\rho^\alpha_{\textbf{k}_1+\textbf{k}_2}}{\rho^\mu_{\textbf{k}_1+\textbf{k}_2}}\cors{\rho^\beta_{\textbf{k}_1}}{\rho^\mu_{\textbf{k}_1}}\cors{\rho^\gamma_{\textbf{k}_2}}{\rho^\mu_{\textbf{k}_2}}\nonumber.
\end{align}
We assume that the convolution approximation holds throughout this work. While, technically, this is another approximation of MCT, it is well-documented that it has little influence on the MCT predictions in fragile liquids \cite{pihlajamaa2023unveiling, bosse1982mode, barrat1989liquid, sciortino2001debye, kob2002quantitative, ayadim2011mode, coslovich2013static, pihlajamaa2023emergent} and thus we do not treat it further in this work.

The operator projecting onto the space of orthogonalized doublets is
\begin{equation}
    \pp_2 = \frac{1}{4}\sum_{\textbf{k}_1\ldots\textbf{k}_4}\ket{Q^{\alpha\beta}_{\textbf{k}_1\textbf{k}_2}}\cors{Q^{\alpha\beta}_{\textbf{k}_1\textbf{k}_2}}{Q^{\gamma\delta}_{\textbf{k}_3\textbf{k}_4}}^{-1}\bra{Q^{\gamma\delta}_{\textbf{k}_3\textbf{k}_4}}.
\end{equation}
Together with Eq.~\eqref{eq:R_doublets}, this implies that $R^{\alpha}_\textbf{k} = \pp_2R^{\alpha}_\textbf{k}$. Without approximation, we may therefore write the memory kernel as
\begin{equation}
    M^{\alpha\beta}(k,t) \approx \frac{1}{k^2D^\beta N^\beta}\cort{R_\textbf{k}^\alpha}{\pp_2e^{\varOmega t}\pp_2}{R_\textbf{k}^\beta}.
\end{equation}
Carrying out the projections yields a double wave-vector integral of the off-diagonal four-point density correlation function, with static prefactors called vertices $V^{\mu\nu\alpha}(\textbf{q}_1, \textbf{q}_2, \textbf{k})$. The result of this projection is
\begin{align}\label{eq:Moff}
&\left(\textbf{M}_\mathrm{offdiag}\right)^{\alpha\beta}(k,t)=\frac{D^\alpha\rho^2}{4x^\beta k^2 N^3} \sum_{\textbf{q}\textbf{q}'} V^{\mu'\nu'\alpha}(\textbf{q}', \textbf{k}-\textbf{q}', \textbf{k})\\&\nonumber\times\corte{Q_2^{\mu'\nu'}(\textbf{q}', \textbf{k}-\textbf{q}')}{Q_2^{\mu\nu}(\textbf{q}, \textbf{k}-\textbf{q})}V^{\mu\nu\beta}(\textbf{q}, \textbf{k}-\textbf{q}, \textbf{k}).
\end{align}
The vertices can be approximated as
\begin{align}
    &V^{\mu\nu\alpha}(\textbf{q}_1, \textbf{q}_2, \textbf{k})\\ &= \frac{N}{2D^\alpha\rho}\sum_{\textbf{k}_1\textbf{k}_2} \cors{Q_2^{\mu\nu}(\textbf{q}_1,\textbf{q}_2)}{Q_2^{\mu'\nu'}(\textbf{k}_1,\textbf{k}_2)}^{-1}\cors{\rk{1}^{\mu'}\rk{2}^{\nu'}}{R_\textbf{k}^\alpha}\nonumber\\\nonumber
    &\approx (\textbf{q}_1\cdot\textbf{k})c^{\mu\alpha}(q_1)\delta^{\nu\alpha} + ((\textbf{k}-\textbf{q}_1)\cdot\textbf{k})c^{\nu\alpha}(\textbf{k}-\textbf{q}_1)\delta^{\mu\alpha},
\end{align}
where we have diagonalized and factorized the normalization of $\pp_2$, again used the convolution approximation, and introduced the direct two-point correlation function $\rho c^{\mu\nu}(k) = \delta^{\mu\nu}/x^\mu - \left(\textbf{S}^{-1}(k)\right)^{\mu\nu}$. 
Equation \eqref{eq:Moff} is an approximation of $\textbf{M}_\varOmega(k,t)$ due to the approximations applied to the inverse static four-point function in the vertices. The resulting memory kernel $\textbf{M}_\mathrm{offdiag}(k,t)$ now contains a dynamic off-diagonal four-point correlation function. By subsequently diagonalizing and factorizing this function, we arrive at the final MCT memory kernel $\textbf{M}_{\mathrm{MCT}}$. 

The above derivation can be summarized with the following consecutive approximations of the memory kernel: 
\begin{widetext}
\begin{align}
    \left(\textbf{M}_\varOmega\right)^{\alpha\beta}(k,t) =& \frac{1}{k^2N^\beta D^\beta}\cort{R_\textbf{k}^\alpha}{e^{\varOmega t}}{R_\textbf{k}^\beta}; \\
    \left(\textbf{M}_\mathrm{offdiag}\right)^{\alpha\beta}(k,t)=& \frac{D^\alpha\rho^2}{4x^\beta k^2 N^3} \sum_{\textbf{q}\textbf{q}'} V^{\mu'\nu'\alpha}(\textbf{q}', \textbf{k}-\textbf{q}', \textbf{k})\corte{Q_2^{\mu'\nu'}(\textbf{q}', \textbf{k}-\textbf{q}')}{Q_2^{\mu\nu}(\textbf{q}, \textbf{k}-\textbf{q})}V^{\mu\nu\beta}(\textbf{q}, \textbf{k}-\textbf{q}, \textbf{k});\\
    \left(\textbf{M}_\mathrm{mct}\right)^{\alpha\beta}(k,t)=& \frac{D^\alpha\rho^2}{2x^\beta k^2 N^3} \sum_{\textbf{q}} V^{\mu'\nu'\alpha}(\textbf{q}, \textbf{k}-\textbf{q}, \textbf{k}) 
    \corte{\rho^{\mu'}_{\textbf{q}}}{\rho^{\mu}_{\textbf{q}}}
    \corte{\rho^{\nu'}_{\textbf{k}-\textbf{q}}}{\rho^{\nu}_{\textbf{k}-\textbf{q}}}
    V^{\mu\nu\beta}(\textbf{q}, \textbf{k}-\textbf{q}, \textbf{k}).\label{eq:mct}
\end{align}
\end{widetext}
\begin{figure*}[ht]
    \centering
    \includegraphics[width=\textwidth]{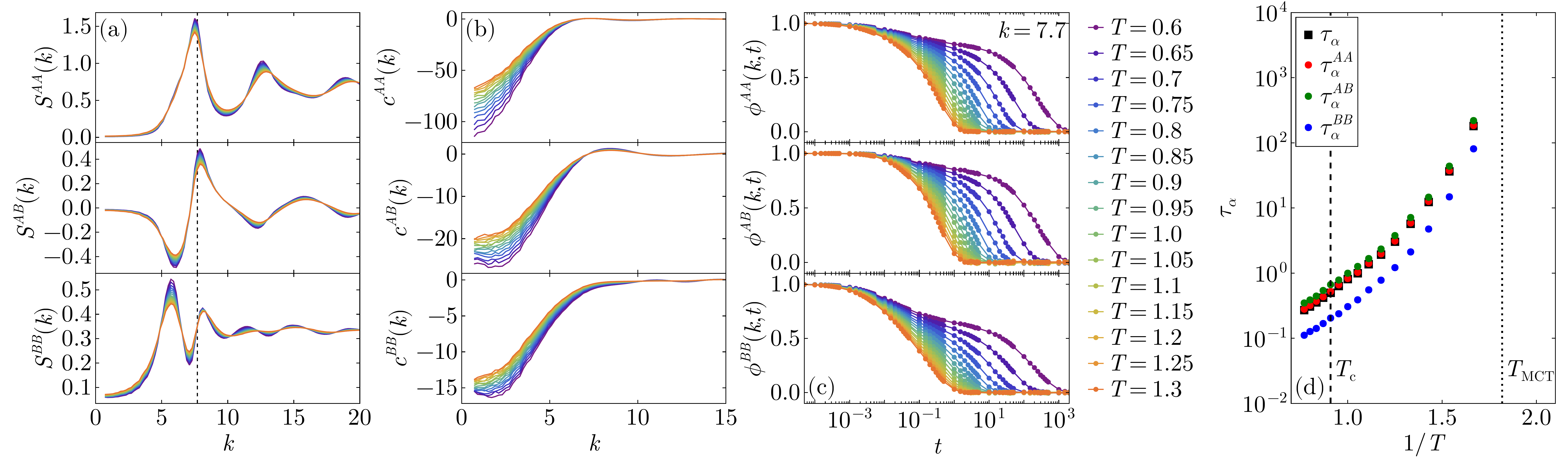}
    \caption{Statics and dynamics of the 2:1 KALJ model. In panels (a) and (b), we show the static structure factor $\textbf{S}(k)$ and the direct correlation functions $\textbf{c}(k)$ for each of the species combinations as a function of $k$. In panel (a), the peak of the total structure factor is indicated with a black dashed line. Panel (c) shows the normalized intermediate scattering function $\phi^{\alpha\beta}(k,t) = F^{\alpha\beta}(k,t)/S^{\alpha\beta}(k)$ as a function of $t$ at the peak of the total structure factor. Panel (d) shows the associated relaxation time, obtained by requiring that $\phi(k, \tau_\alpha) = 1/e$. The vertical dashed lines indicate the ideal glass transition temperature $T_\mathrm{c}$ as predicted by self-consistent mode-coupling theory calculations, and the power-law to Arrhenius crossover temperature $T_\mathrm{MCT}$ obtained by a power-law fit performed by \citet{ortlieb2023probing}. }
    \label{fig:fig1}
\end{figure*}
In the following sections, we compute and compare these memory kernels to investigate the effect of these approximations in supercooled liquids. We shall evaluate the different kernels with the dynamical correlation functions obtained from simulations. This should be contrasted to the standard self-consistent technique, where the MCT memory kernel is evaluated as a bi-linear functional of the intermediate scattering function. Equations \eqref{eq:mct} and \eqref{eq:eom} are then solved self-consistently. To avoid any confusion in the remainder of this paper, we will always denote this procedure as scMCT, referring to a self-consistent iteration with only static simulation input. We emphasize that this is how MCT is usually employed.

We note that, historically, it has been argued that the factorization approximation should be performed simultaneously with the $\left[\exp(\qq \varOmega \qq' \qq t) \to \exp(\varOmega t)\right]$-step (see e.g.\ \citet{gotze2009complex}, p.\ 183). The reason is the following: Clearly, one cannot factorize before changing the dynamics because $\cort{\rho_\textbf{q}^\alpha}{\exp{(\qq \varOmega \qq' \qq t)}}{\rho_\textbf{q}^\beta}$ vanishes by the construction of $\qq$. Conversely, if the factorization is performed after the change of dynamics, G\"otze argues, one neglects contributions proportional to $\corte{\rho_{\textbf{q}_1}^{\mu'} \rho_{\textbf{k}-\textbf{q}_1}^{\nu'}\pp}{\pp\rho_{\textbf{q}_2}^\mu\rho_{\textbf{k}-\textbf{q}_2}^\nu}$.  
However, by using orthogonalized doublets in this work, we have circumvented this issue because $\corte{Q_2^{\mu'\nu'}(\textbf{q}', \textbf{k}-\textbf{q}')\pp}{\pp Q_2^{\mu\nu}(\textbf{q}, \textbf{k}-\textbf{q})}=0$  by construction. In section \ref{sec:results}, we show the degree to which the orthogonalization of the doublets impacts the resulting intermediate scattering function.

Additionally, to derive the theory one may also place the $\left[\exp(\qq \varOmega \qq' \qq t) \to \exp(\varOmega t)\right]$-step between the doublet projection and the factorization approximations. This approach is, however, much harder to follow numerically since computing the quantity $\left<Q_2^{\mu'\nu'}(\textbf{q}', \textbf{k}-\textbf{q}')|e^{\qq\varOmega\qq'\qq t}|Q_2^{\mu\nu}(\textbf{q}, \textbf{k}-\textbf{q})\right>$ for all $\textbf{q}$ and $\textbf{q}'$ is intractable with current methods. 

\section{Simulation model}\label{sec:methods}

As our model supercooled liquid, we use the 2:1 Kob-Andersen Lennard-Jones (KALJ) mixture introduced in \citet{ingebrigtsen2019crystallization}. They have shown that this mixture is more stable against crystallization than the original 4:1 version \cite{ingebrigtsen2019crystallization}. Additionally, the more equal compositional ratio makes it easier in our case to measure memory kernels of the minority component with sufficient statistical accuracy. We simulate a cubic system of $N$ Brownian particles at bulk density $\rho$ using the LAMMPS software \cite{thompson2022lammps}. The mixture consists of a large and a small component, respectively labeled A and B, of which the Lennard-Jones interaction parameters are identical to that of the standard Kob-Andersen mixture, \textit{i.e.}, $\sigma^{\mathrm{AA}}= 1.0$, $\sigma^{\mathrm{BB}}=0.88$, $\sigma^{\mathrm{AB}}=0.8$, $\epsilon^{\mathrm{AA}}=1.0$, $\epsilon^{\mathrm{BB}}=0.5$, and $\epsilon^{\mathrm{AB}}=1.5$. The potentials are cut-off and shifted at $r_{ij}/\sigma^{\alpha\beta}=2.5$. Throughout this work, we present all quantities in units of $\epsilon^{\mathrm{AA}}$ and $\sigma^{\mathrm{AA}}$. We prepare equilibrated samples with $N=1000$ particles at $\rho = 1.4$ in the NVT ensemble using molecular dynamics with a Nosé-Hoover thermostat. Subsequently, we apply the Euler-Maruyama algorithm for the Brownian production runs lasting $10^8$ time steps with time step $\delta t = 5\times10^{-5}$. We repeat this procedure for different temperatures to vary the degree of supercooling. At each temperature, we generate 400 independent simulations. In this work, we consider only state points in the power-law regime $T>T_\mathrm{MCT}$, where $T_\mathrm{MCT}=0.55$ as determined by \citet{ortlieb2023probing} from a power-law fit to their simulation data. 

From the generated data sets, we extract the static and dynamic correlation functions required to compute the memory kernels presented in the previous section. Each of these functions is computed directly in $k$-space, averaging over all possible wave vectors that satisfy $k\in[k_*-\Delta k/2, k_*+\Delta k/2]$, with $\Delta k = 0.2$, and $k_*$ the target wave number. A subset of these functions is presented in Fig.~\ref{fig:fig1}. Specifically, we show the partial static structure factors and direct correlation functions for different temperatures. Upon supercooling, these functions show only very minor quantitative changes. 
The peak of the total structure factor $S(k) = \sum_{\alpha\beta}S^{\alpha\beta}(k)$ is located at $k=7.7$, which is indicated with a dashed vertical line in panel (a). In panels (c) and (d), we present the normalized intermediate scattering functions and the associated relaxation times, which show that our temperature range encompasses relaxation times that vary by three orders of magnitude. Since we are dealing with species-resolved quantities, there are different ways of normalizing the dynamical functions. Specifically, in Fig.~\ref{fig:fig1}(c), we have used the normalization $\phi^{\alpha\beta}(k,t) = F^{\alpha\beta}(k,t)/S^{\alpha\beta}(k)$. While this is perhaps the easiest to interpret, it may give numerical issues for the cross-component terms. Therefore, we henceforth use the more principled \cite{gotze2003effect} normalization $f^{\alpha\beta}(k,t) = \left(\textbf{S}^{-\frac{1}{2}}(k)\textbf{F}(k,t)\textbf{S}^{-\frac{1}{2}}(k)\right)^{\alpha\beta}$, where $\textbf{S}^{-\frac{1}{2}}(k)$ is defined by $\textbf{S}^{-\frac{1}{2}}(k)\textbf{S}(k)\textbf{S}^{-\frac{1}{2}}(k)=\textbf{I}$. This normalization ensures that $\textbf{f}$ is a positive definite, symmetric matrix. The structural relaxation times $\tau_\alpha$ in panel (d) are determined by requiring that the normalized intermediate scattering function has decayed to a value of $1/e$. The data show that the BB-relaxation time is shorter than the others by half an order of magnitude. This, however, is slightly misleading because the peak of $S^\mathrm{BB}(k)$ is not close to $k=7.7$, where $\tau_\alpha^\mathrm{BB}$ is measured. The partial relaxation times measured at the peaks of the corresponding partial structure factors are all similar to the overall relaxation time.

\begin{figure}[ht]
    \centering
    \includegraphics[width=\linewidth]{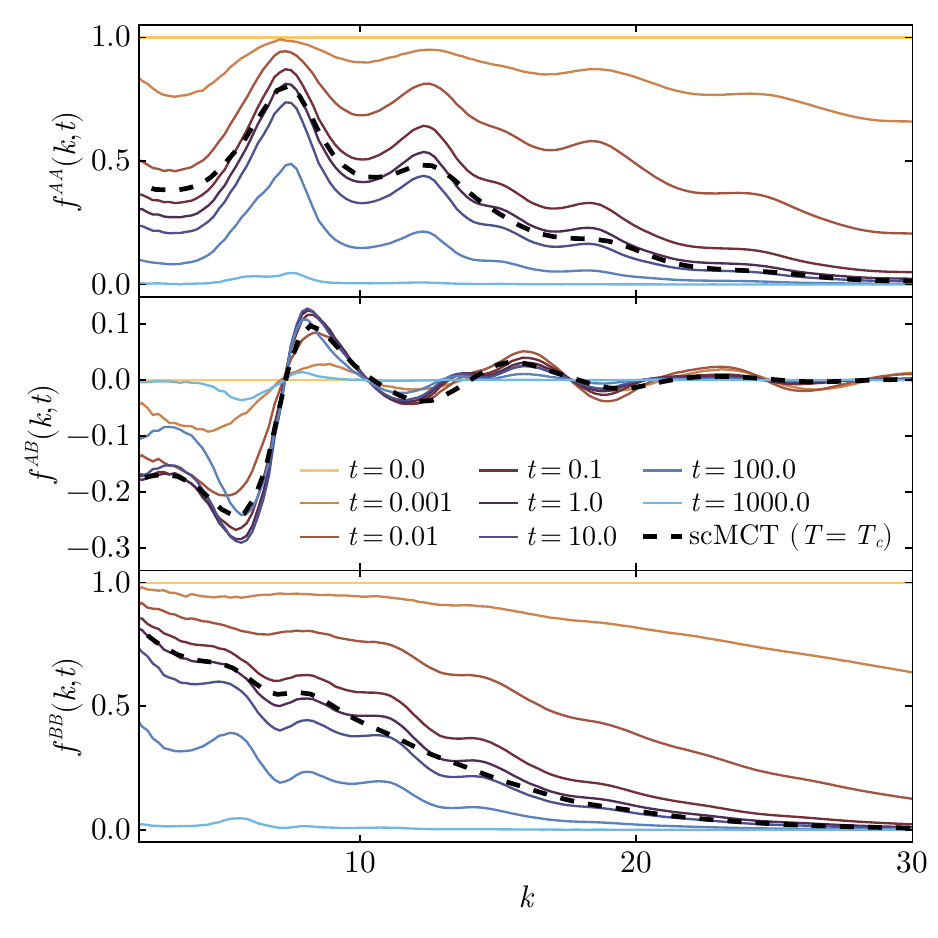}
    \caption{Wave-vector dependence of the normalized intermediate scattering function $\textbf{f}(k,t) = \textbf{S}^{-1/2}(k)\textbf{F}(k,t)\textbf{S}^{-1/2}(k)$ at different $t$, together with the critical nonergodicity parameter predicted by mode-coupling theory. Each panel shows the simulation data for different components. The simulation data (full lines) are generated at $T=0.6$, and the nonergodicity parameter of scMCT is obtained from $T_\mathrm{c}=1.1$ (with the structure factor at the same temperature as input).}
    \label{fig:fig2}
\end{figure}

\begin{figure*}[ht]
    \centering
    \includegraphics[width=\textwidth]{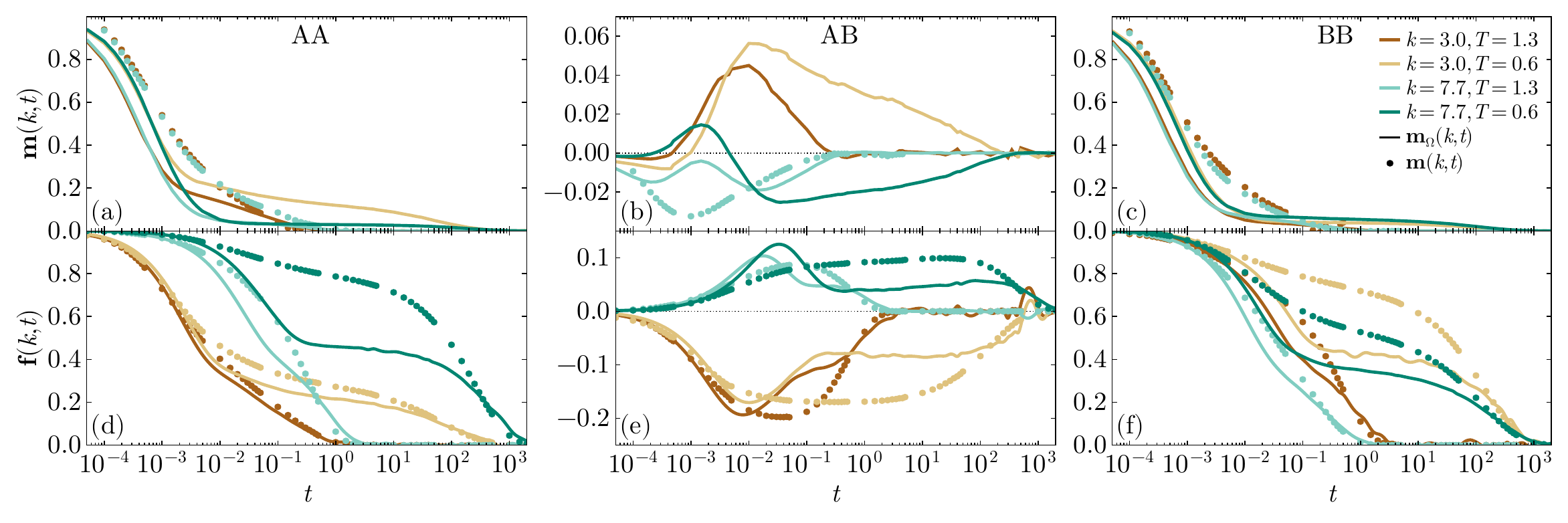}
    \caption{Normalized kernel with unprojected dynamics $\textbf{m}_\varOmega(k,t) = (\textbf{M}_\varOmega(k, 0))^{-1/2}\textbf{M}_\varOmega(k,t)(\textbf{M}_\varOmega(k,0))^{-1/2}$ as a function of $t$ for values of $k$ and $T$ as indicated in the legend. Panels (a--c) show the measured memory kernel, and panels (d--e) show the corresponding intermediate scattering function $\textbf{f}_\varOmega(k,t) = (\textbf{F}_\varOmega(k, 0))^{-1/2}\textbf{F}_\varOmega(k,t)(\textbf{F}_\varOmega(k,0))^{-1/2}$, obtained by solving the generalized Langevin equation. Panels (a, d), (b, c), and (c, f) show the AA, AB, and BB contributions respectively. The true results are indicated with circular markers, obtained either through direct measurement in the case of $\textbf{f}$ or by the solution of the Volterra integral equation in the case if $\textbf{m}(k,t)$. In cases where the latter did not converge to an acceptable result, the data is not shown.}
    \label{fig:fig3}
\end{figure*}

For this mixture, self-consistent mode-coupling theory (scMCT) predicts an ideal glass transition at $T_{\mathrm{c}}=1.1$. This value is obtained using the standard integration techniques \cite{pihlajamaa2023modecouplingtheory, fuchs1991comments} with only the static structure factors from simulation as input using an equidistant wave-vector grid with $N_k=200$ points between $k=0.2$ and $k=78.8$. The non-ergodicity parameter $\textbf{f}(k,t\to\infty)$ predicted by MCT at the critical point is shown together with the simulation data for $\textbf{f}(k,t)$ as a function of $k$ at different $t$ in Fig.~\ref{fig:fig2}. We show simulation data at $T=0.6$ as this is our closest point to the dynamic crossover point $T_{\mathrm{MCT}}=0.55$. 

The data show that the predicted non-ergodicity parameter (dashed, black line in Fig.~\ref{fig:fig2}) gives excellent qualitative and semi-quantitative agreement with the measured correlator at $t\approx1$, which corresponds to the onset of the plateau of $\textbf{f}(k,t)$ as shown in Fig.~\ref{fig:fig1}(c). From the data, it is clear that the shape of the AA- and AB-components of $k$-dependence of the plateau height of $\textbf{f}(k,t)$ (the Debye-Waller factors) are much more strongly governed by the structure factor than the BB-function, as is the case with the standard 4:1 mixture. Interestingly the cross-component correlator shows non-monotonic behavior when normalized as $\textbf{f}$: it starts at $f^{\mathrm{AB}}(k,0)=0$ (as imposed by the normalization), then increases, reaching a maximum when the original function plateaus, to subsequently decorrelate. 


\section{Results}\label{sec:results}

In this section, we compute the approximate memory kernels and the intermediate scattering functions that they correspond to [cf.~Eq.~\eqref{eq:eom}]. We start with the kernel $\textbf{M}_\varOmega(k,t)$ that evolves with standard instead of orthogonal dynamics. It is computed from simulations by direct measurement of the correlation function of the fluctuating force. In particular, we define $\ket{h^\alpha_\textbf{k}} = \varOmega\ket{\rho_\textbf{k}^\alpha} = D^\alpha\sum_{j=1}^{N^\alpha}\left[i\beta (\textbf{k}\cdot\textbf{F}_j^\alpha)-k^2\right]e^{i(\textbf{k}\cdot\textbf{r}^\alpha_j)}$, and thus 
\begin{align}\label{eq:RR}
    \nonumber\corte{R^\alpha_\textbf{k}}{R^\beta_\textbf{k}}/N &= \corte{h^\alpha_\textbf{k}}{h^\beta_\textbf{k}}/N -  \nonumber (\textbf{H}(k)\textbf{S}^{-1}(k)\textbf{H}(k,t))^{\alpha\beta}\nonumber\\  &- (\textbf{H}(k,t)\textbf{S}^{-1}(k)\textbf{H}(k))^{\alpha\beta} \\ &+ (\textbf{H}(k)\textbf{S}^{-1}(k)\textbf{F}(k,t)\textbf{S}^{-1}(k)\textbf{H}(k))^{\alpha\beta}\nonumber,
\end{align}
where $\textbf{H}^{\alpha\beta}(k,t) = - \corte{\rho^\alpha_\textbf{k}}{h^\beta_\textbf{k}}/N$. This is the time-dependent version of the static diffusion matrix $\textbf{H}(k)$. Each correlation function in the expansion of $\corte{R^\alpha_\textbf{k}}{R^\beta_\textbf{k}}$ can be measured from the simulation trajectories since only particle positions and forces are needed, allowing us to readily compute $\textbf{M}_\varOmega(k,t)$. Substituting this memory kernel in Eq.~\eqref{eq:eom} and numerically solving the resulting integro-differential equation \cite{pihlajamaa2023modecouplingtheory}, we find the intermediate scattering function $\textbf{F}_\varOmega(k,t)$ as predicted by $\textbf{M}_\varOmega(k,t)$. This allows us to establish the impact of the approximation not only on the memory kernel itself, but also on its associated dynamical observable. We show the obtained results in Fig.~\ref{fig:fig3} for two different temperatures and wavelengths. Both the memory kernel and the resulting intermediate scattering function are normalized such that they are symmetric and positive definite. 

For reference, we have also attempted to obtain the exact memory kernel with the reduced dynamics $\textbf{M}(k,t)$ by solving the Volterra integral equation,  Eq.~\eqref{eq:volterra}. Unfortunately, our data was insufficiently accurate to avoid numerical instabilities at low temperatures. For the diagonal matrix entries at high temperatures, we have indicated the exact memory kernels in the top panels of the figure with circular markers. The markers in the bottom panels correspond to direct measurements of the intermediate scattering function. 

Our results of the diagonal components of $M_\varOmega^{\alpha\beta}(k,t)$ and the associated $F_\varOmega^{\alpha\beta}(k,t)$ show that simplifying the evolution operator decreases the amount of memory present in the system significantly, resulting in clearly reduced plateau heights, especially near the peak of the structure factor. While less pronounced, this trend persists at higher temperatures, lower wave numbers, and in single-component systems \cite{pihlajamaa2023unveiling}. The off-diagonal entries $F_\varOmega^{\mathrm{AB}}(k,t)$ show clear qualitative deviations from the exact intermediate scattering function for all temperatures and wave vectors shown. Despite the qualitative and quantitative deviations, the kernels and correlators do show final decay on timescales roughly consistent with the true ones. While noteworthy, this is not very surprising due to the use of the simulated intermediate scattering function in the evaluation of Eq.~\eqref{eq:RR}.  

\begin{figure}[t]
    \centering
    \includegraphics[width=\linewidth]{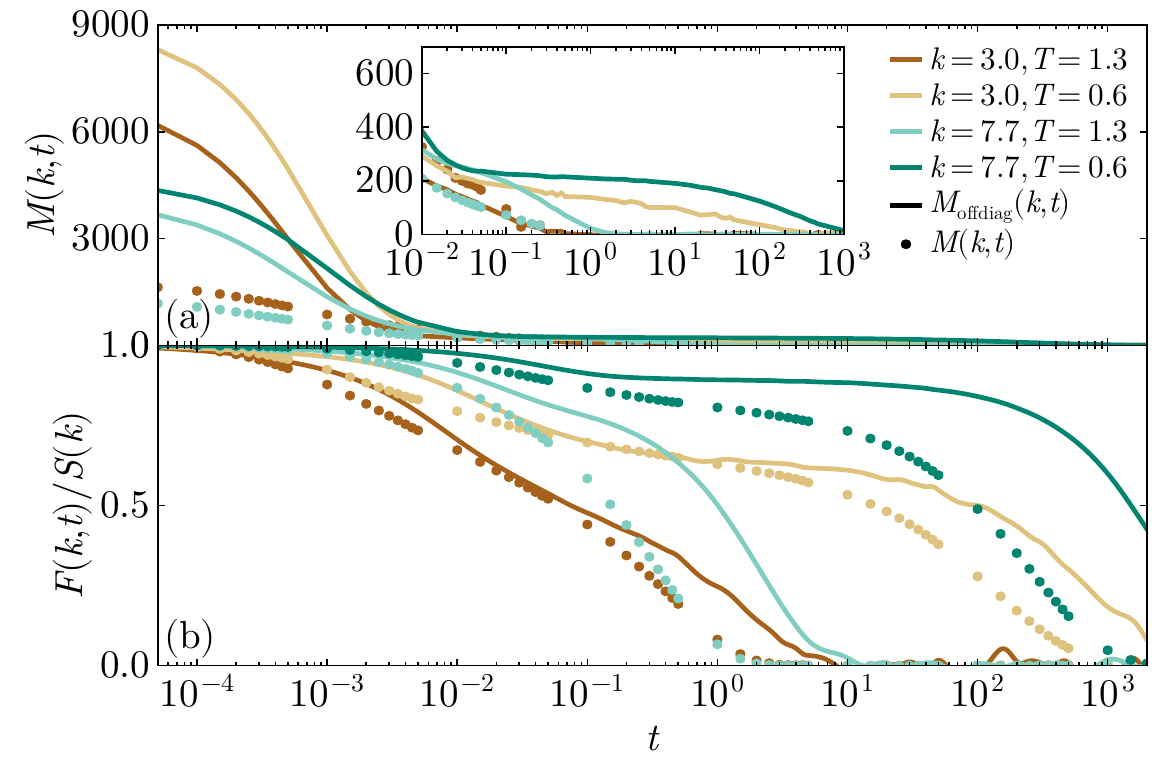}
    \caption{Full memory kernel $M(k,t) = \sum_{\alpha\beta}M^{\alpha\beta}(k,t)$  (a) and normalized full intermediate scattering function $F(k,t)/S(k)$ (b) after the projection on orthogonalized doublets compared with the exact results. The solid lines correspond to the doublet projected functions and the markers correspond to the `exact' results.  }
    \label{fig:fig4}
\end{figure}

We continue by discussing the next approximation: the projection on density doublets. The off-diagonal memory kernel $\textbf{M}_\mathrm{offdiag}(k,t)$, that is, the result of the concomitant approximation of the vertices, is evaluated by the same procedure as outlined in \citet{pihlajamaa2023unveiling} (adapted for mixtures). The wave number cut-off of the sum was chosen at $k=60.0$ to ensure convergence. In contrast to our earlier work \cite{pihlajamaa2023unveiling}, here we perform the orthogonalization with respect to the singlet space explicitly, see Eq.~\eqref{eq:Q2}. The results are visualized in Fig.~\ref{fig:fig4} for the total memory kernel $M(k,t)$ and density correlator $F(k,t)$, which are defined as $M(k,t) = \sum_{\alpha\beta}M^{\alpha\beta}(k,t)$ and 
$F(k,t) = \sum_{\alpha\beta}F^{\alpha\beta}(k,t)$.

Consistent with our earlier work, we find that while the kernel resulting from the previous approximation underestimates the degree of memory, $\textbf{M}_\mathrm{offdiag}(k,t)$ overestimates it. This is also the case for all individual matrix components AA, AB, and BB of the doublet projected kernel. The overestimation is strongest at short times \cite{gotze2009complex}, and around the peak of the structure factor at long times. We find no clear deviations from these observations when varying the temperature. In contrast to neglecting the orthogonal projectors in the evolution of the memory kernel, here, the relaxation time of the intermediate scattering function is significantly affected by the approximation, as can be observed by comparing the final decay time of the lines and markers in the bottom panel of Fig.~\ref{fig:fig4}. Moreover, this overestimation of the relaxation time increases as the temperature is decreased.

The figure shows results only of $\textbf{M}_\mathrm{offdiag}$, which is projected on the doublet space orthogonal to the density singlet space. We have repeated the calculation without the orthogonalization step (as performed in Ref.~\cite{pihlajamaa2023unveiling}) and have found that the resulting kernel slightly increases the overestimation of the memory kernel compared to $\textbf{M}_\mathrm{offdiag}$, which is the result of a projection on the full doublet space. This modest difference is most significant around the peak of the structure factor. It however does not affect the results qualitatively. We will later show the impact of this at a low temperature in Fig.~\ref{fig:fig7}. 

\begin{figure}[t]
    \centering
    \includegraphics[width=\linewidth]{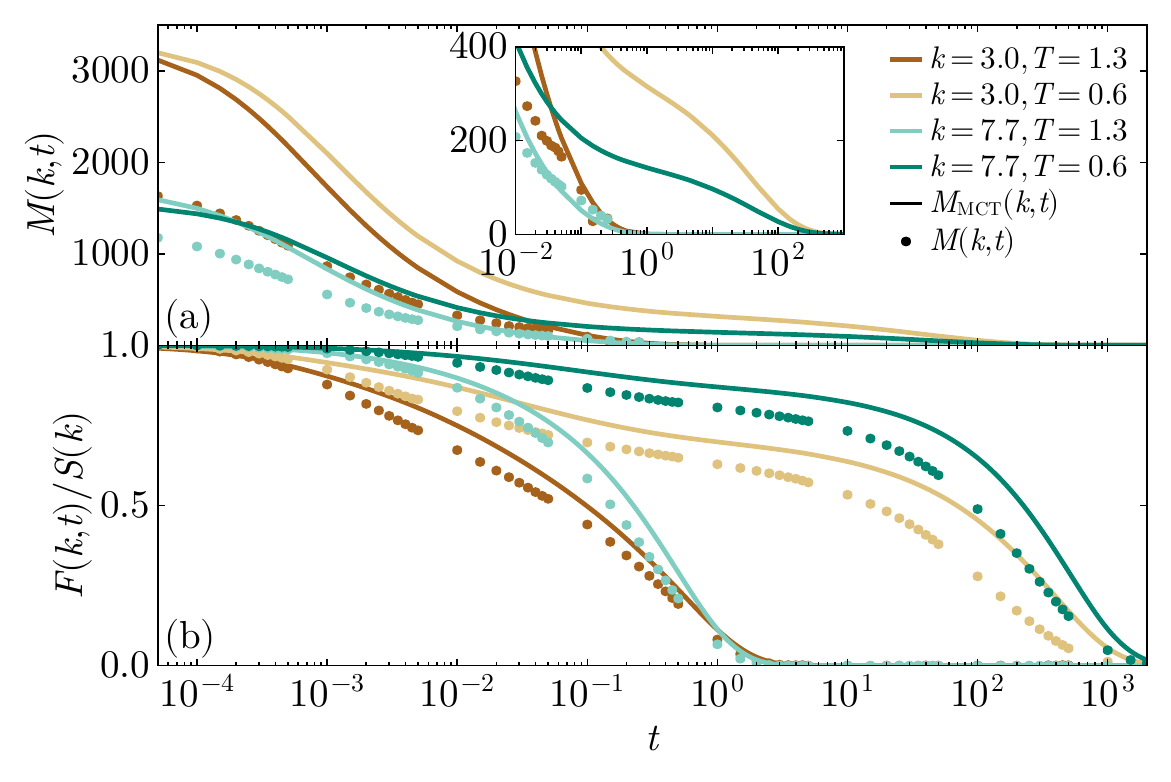}
    \caption{Full memory kernel $M(k,t) = \sum_{\alpha\beta}M^{\alpha\beta}(k,t)$  (a) and normalized full intermediate scattering function $F(k,t)/S(k)$ (b) according to the mode-coupling theory compared with the exact results. The mode-coupling memory kernel is evaluated using time-dependent correlation functions measured in the simulations and not solved self-consistently. The solid lines correspond to the MCT predictions and the markers to the exact ones. }
    \label{fig:fig5}
\end{figure}

Lastly, we show the memory kernel and intermediate scattering function from the non-self-consistent mode-coupling kernel in Fig.~\ref{fig:fig5}. It is important to note that the data are not obtained by self-consistent iteration, which is the standard way that the mode-coupling theory is applied. Instead, our data are determined from the direct computation of Eq.~\eqref{eq:mct}, with the intermediate scattering function from simulations as input. The data is presented as a function of $t$ for different temperatures and wave numbers.

\begin{figure*}[ht]
    \centering
    \includegraphics[width=\textwidth]{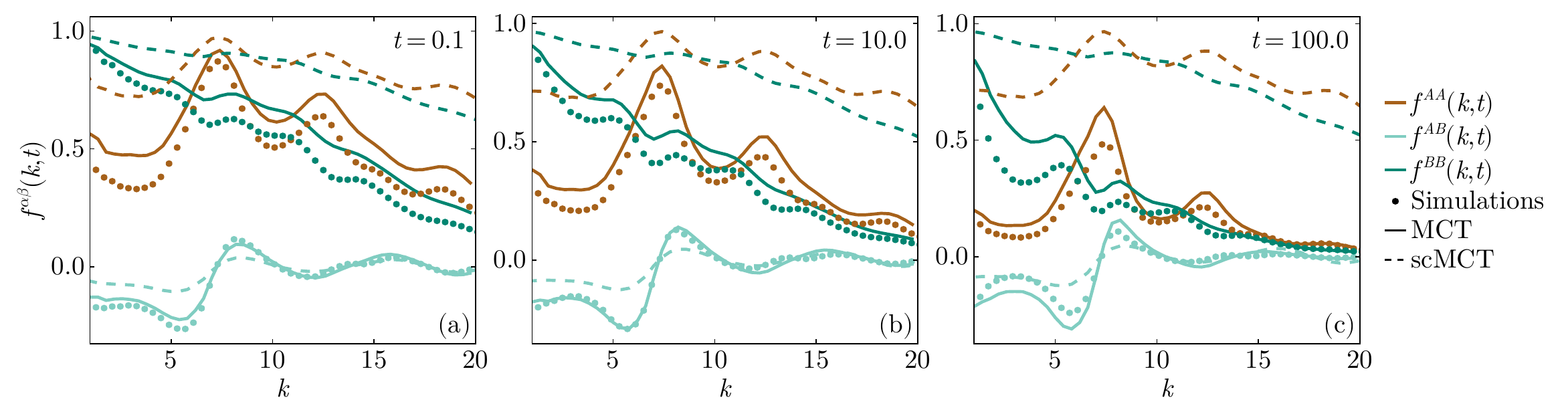}
    \caption{Comparison between the $k$-dependence of the normalized intermediate scattering function obtained directly from simulations, from the mode-coupling memory kernel, and from a self-consistent solution of MCT, respectively indicated with symbols, full and dashed lines. We show the intermediate scattering function at three different time scales corresponding to (a) the relaxation towards the plateau, (b) relaxation away from the plateau, and (c) final structural relaxation, at $T=0.6$. The different colors correspond to the different matrix entries of the partial intermediate scattering function.}
    \label{fig:fig6}
\end{figure*}

The results in Fig.~\ref{fig:fig5} show that the mode-coupling kernel also overestimates the memory of the supercooled liquid, but that it does so to a much smaller extent than the off-diagonal kernel. This is consistent across the temperatures and wavelengths studied, where the deviation between MCT and simulation only moderately increases as the degree of supercooling is increased. This slight overestimation seems to be partly caused by an excess of memory at short and intermediate times as indicated by the early deviations of the correlators. This overestimation of the short-time memory can be seen by comparing the MCT kernel (lines) to the exact memory kernel (markers) for the highest temperatures shown in Fig.~\ref{fig:fig5}(a). At lower temperatures and later times, the excess memory of MCT increases slightly, as observed by the increased overestimation of $F(k,t)$ [Fig.~\ref{fig:fig5}(b)]. This indicates that in deeply supercooled conditions long-time errors are introduced in the memory kernel. These are perhaps the same that cause the failure of MCT to capture the dynamic crossover.

\begin{figure*}[t]
    \centering
    \includegraphics[width=\linewidth]{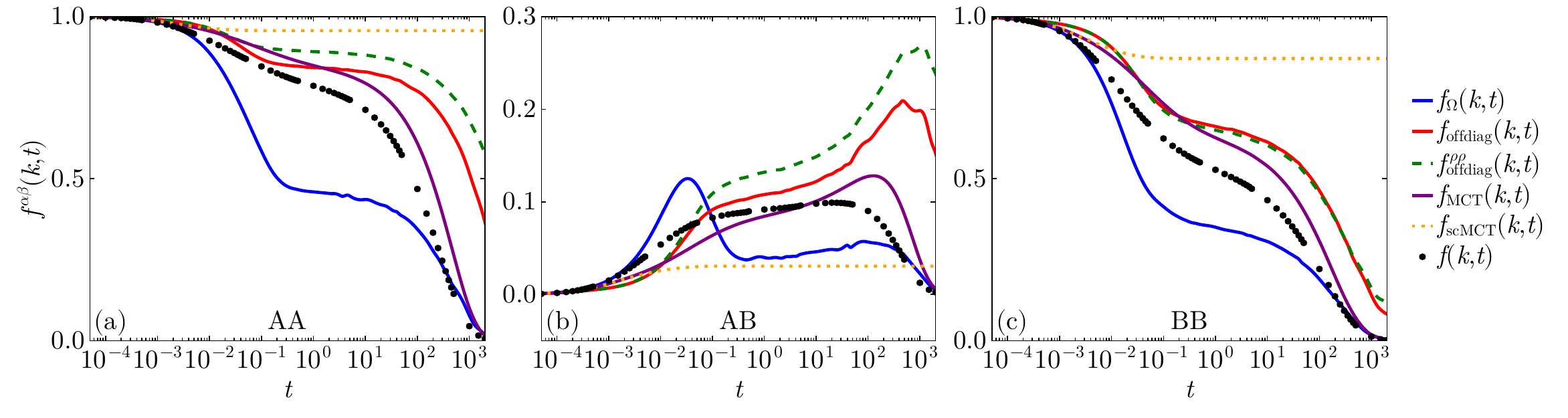}  
    \caption{Normalized intermediate scattering functions as a function of $t$ for $k=7.7$ and $T=0.6$ obtained from the different approximations of the memory kernel. Panels (a--c) show the AA, AB, and BB matrix entries respectively. The true (simulation) results are indicated with black circles. The superscript $\rho\rho$ in the legend indicates that the corresponding off-diagonal kernel is computed without proper orthogonalization with respect to the space of density singlets.}
    \label{fig:fig7}
\end{figure*}

Mode-coupling theory distinguishes three different time scales that correspond to the relaxation (1) towards the plateau of $\textbf{F}(k,t)$ (early $\beta$-decay), (2) away from the plateau (the Von Schweidler-law), and (3) final structural decorrelation ($\alpha$-decay). To understand the accuracy of the predictions of MCT in these three regimes, we compare simulation data with the intermediate scattering functions from the MCT memory kernel, solved both self-consistently and with the scattering function from simulations as input. We show the result as a function of $k$ for $T=0.6$ in Fig.~\ref{fig:fig6}. At this temperature, the three regimes roughly coincide with $t=0.1$, $t=10$, and $t=100$ [see also Fig.~\ref{fig:fig5}(b)]. 

Our results demonstrate a strong qualitative agreement between the simulation results and the correlation function obtained from the MCT memory kernel across the entire range of wavelengths studied, with the best agreement observed in the cross-component scattering function. 
While the self-consistent solution retains some of the key qualitative features, it performs much worse quantitatively. 
Examining different time regimes (comparing different panels in Fig.~\ref{fig:fig6}), we find that the most striking deviations of the mode-coupling theory originate at early times, where it overestimates the relaxation time for the AA and BB components for all $k$. This confirms that MCT overestimates the amount of memory at short times.
For the AA component, this disagreement is least notable at the first peak of the structure factor, where the results give very good agreement. 
As $t$ is increased, the agreement between theory and simulation seems to increase overall (see also Fig.~\ref{fig:fig5}(b)). An exception is found at the first peak of the structure factor, where MCT progressively overestimates the relaxation time as $t$ is increased. This indicates that MCT overestimates the effect of caging.  Additionally, for the AB and BB components, the deviation grows for small $k$ as $t$ is increased into the structural relaxation regime. 

We summarize our results in Fig.~\ref{fig:fig7}, by showing the normalized intermediate scattering function $\textbf{f}(k,t)$ after each approximation at the $T=0.6$ and $k=7.7$ together with $\textbf{f}(k,t)$ measured directly from simulations. In addition to the approximations described above, we also show $f^{\rho\rho}_\mathrm{offdiag}(k,t)$, which is computed with the memory kernel $M^{\rho\rho}_\mathrm{offdiag}(k,t)$, the result of a doublet projection step that is not orthogonalized with respect the density singlets. 

From the results, it is clear that neglecting the projectors in the time evolution exponent of the memory kernel, that is, going from $\textbf{f}(k,t)$ to $\textbf{f}_\varOmega(k,t)$ results in an initial relaxation that is too fast, yielding a plateau that is too low. The final relaxation time, however, is not affected much. The next step, the projection on doublets and the associated diagonalization/factorization of the inverse static four-point structure factor in the vertices, is shown to undo all of the effects of the previous approximation: both the initial and final relaxation time of $\textbf{f}_\mathrm{offdiag}(k,t)$ are much slower than that of the simulation data. Whether the doublet projector is orthogonalized with respect to the singlet space makes little qualitative difference. Both overestimate the final relaxation time by more than an order of magnitude. Next, diagonalizing and factorizing the dynamic four-point correlation function, that is, going to $\textbf{f}_\mathrm{MCT}$, almost entirely cancels the effects of the same approximation on the inverse function done in the previous step. In this sense, the theory relies on a large cancellation-of-errors effect, showing that it is essential to treat the dynamics and statics on an equal footing within this kinetic theory. The function $\textbf{f}_\mathrm{MCT}(k,t)$ shows good agreement with $\textbf{f}(k,t)$ even at temperatures near the mode-coupling temperature $T_\mathrm{MCT}$, only slightly overestimating the plateau height. However, if the theory is solved self-consistently [$\textbf{f}_\mathrm{scMCT}(k,t)$], this slight overestimation is expanded drastically. We can attribute this to the nonlinear feedback loop of the self-consistent iteration, resulting in a prediction of a deep glass state. The above observations are consistent with our earlier work on a single-component liquid \cite{pihlajamaa2023unveiling}.

\section{Discussion and Conclusion}\label{sec:conclusion}

We have tested each of the approximations made to mode-coupling theory in a supercooled binary mixture against Brownian dynamics simulations. While each approximation clearly affects the predicted dynamics, we find that, overall, there is a cancellation of errors among these approximations at long times. In particular, this cancellation is due to the static diagonalization/factorization approximation in the normalization of the doublet projector and the dynamic diagonalization/factorization approximation of the four-point scattering function. This emphasizes that dynamics and statics should be treated equally in the theory. These findings are also consistent with earlier results for a single-component liquid above the melting point \cite{pihlajamaa2023unveiling}.

Additionally, we find that the final MCT approximation of the memory kernel remains very accurate, even in the supercooled regime. The final discrepancies between the predictions of the theory and experiment arise from the self-consistent manner in which the MCT equations are solved, greatly magnifying the small errors of the memory kernel. While unfortunate in practice, it is also encouraging because it means that even small corrections to the memory functional can drastically improve the predictions of the self-consistently solved theory. This suggests that the current bi-linear form of the memory functional is accurate up to small corrections and should be retained at the lowest order by an improved mode-coupling theory. Furthermore, we find little qualitative change in the validity of the approximations as the liquid is supercooled or if multiple components are introduced \cite{pihlajamaa2023unveiling}. We therefore believe that improving the theory in high-temperature liquid conditions is perhaps sufficient to cure many of MCT's quantitative flaws at supercooled temperatures down to the power-law-to-Arrhenius crossover point. 

Concerning the short-time errors introduced in the theory, our results indicate that these early-time deviations originate from another cancellation of errors between the static diagonalization/factorization and the approximation of the orthogonal dynamics, culminating in a net overestimation of short-time memory. In contrast to the case of Newtonian systems, these short-time contributions cannot be attributed to unaccounted-for current-couplings \cite{bosse1978mode}, because in Brownian systems the doublet projection is an exact procedure, and the density modes are expected to contain all relevant physics.

The second source of error arises at long times, where MCT overestimates the amount of memory around the first peak of the static structure factor of the majority species. Physically, this implies that MCT puts too much weight on the effect of caging because the theory does not include ergodicity-restoring activated processes that allow particles to escape their cages. The associated error grows with increasing time and decreasing temperature, leading to the critical behavior of the theory. We expect that these errors will proliferate at lower temperatures than those studied here (that is, below the dynamic crossover at $T_\mathrm{MCT}$). It has been argued by \citet{cates2006current} that the diagonalization/factorization approximation leading to the bi-linear functional form of the memory kernel breaks down in this regime, due to the distinctly non-Gaussian nature of the density fluctuations corresponding to activated ergodicity-restoring processes. According to their argument, the correct kernel might be closer to a linear functional in the activated regime, resulting in a smearing-out of the critical transition as observed in simulations. Hence, a proper theoretical account of these activated processes likely requires a functional form different from the current bi-linear one to extend the range of validity down to the experimental glass transition. 


\section*{Acknowledgements}
We would like to express our gratitude to Grzegorz Szamel and Corentin C. L. Laudicina for very useful discussions. 
The Dutch Research Council (NWO) is acknowledged for financial support through a Vidi grant (IP and LMCJ).

\section*{References}
\bibliography{aipsamp}

\end{document}